\documentclass[leqno]{article}
\usepackage[english]{babel}
\usepackage[cp850]{inputenc}
\usepackage{amsthm}
\usepackage{amscd}
\usepackage{amssymb}
\usepackage{amsmath}
\usepackage{verbatim}
\usepackage{graphics}
\usepackage{graphicx}
\usepackage{color}

\newtheorem{myexa}{Example}

\newtheorem{mytheo}{Theorem}
\newtheorem{mylem}{Lemma}
\newtheorem{myprop}{Proposition}
\newtheorem*{myrema}{{\em Remark}}

\newtheorem*{myproo}{{\em Proof}}

\definecolor{aggrey}{rgb}{.7,.7,.7}

\begin{document}
\thispagestyle{empty} \vspace*{-1cm}

%\scalebox{0.2}{\includegraphics[width=1250pt,height=200pt]{fim2.pdf}}
%%\hfill \textcolor{aggrey}{STATISTICS 2011 CANADA/ IMST-2011-FIM XX}
%\hfill
%\scalebox{0.5}{\includegraphics[width=55pt,height=90pt]{escudouex.pdf}}
% \vspace*{8cm}
%
%
%
%\large

\begin{center}
{\sc  A Monte Carlo Method to Approximate Conditional Expectations based on a}\\
{\sc Theorem of Besicovitch: Application to Equivariant Estimation}\\ {\sc of the Parameters of the General Half-Normal Distribution}\vspace{1cm}\\
A.G. Nogales$^*$, P. P\'erez$^*$ and P. Monfort$^{**}$\vspace{1cm}\\
Dpto. de Matem\'aticas, Universidad de Extremadura\\
Avda. de Elvas, s/n, 06071--Badajoz, SPAIN.\\
e-mail: nogales@unex.es
\end{center}
\vspace{.4cm}
\begin{quote}
\hspace{\parindent} { {\sc Abstract.}

A natural Monte Carlo method to approximate conditional expectations in a probabilistic framework is justified by a general result inspired on the Besicovitch covering theorem on differentiation of measures. The method is specially useful when densities are not available or are not easy to compute.
The method is illustrated by means of some examples and can also be used in a statistical setting to approximate the conditional expectation given a sufficient statistic, for instance. In fact, it is applied to evaluate the minimum risk equivariant estimator (MRE) of the location parameter of a general half-normal distribution since this estimator is described in terms of a conditional expectation for known values of the location and scale parameters.  For the sake of completeness, an explicit expression of the the minimum risk equivariant estimator of the scale parameter is given. For all we know, these estimators have not been given before in the literature. Simulation studies are realized to compare the behavior of these estimators with that of maximum likelihood and unbiased estimators.}
 \end{quote}
%\end{document}

\vfill
\begin{itemize}
\item[] \hspace*{-1cm} {\em AMS Subject Class.} (2010): {\em
Primary\/} 65C05,  62F10, 62-04. %62E17
\item[] \hspace*{-1cm} {\em Key words and phrases:} Monte Carlo simulation, conditional expectation, general half-normal distribution, equivariant estimation.
%\item[] \hspace*{-1cm} {\em Short running title:} General
%Half-Normal: point estimation of the parameters.
\item[] \hspace*{-1cm} (*) This author has been supported by the spanish {\it Ministerio de Ciencia e Innovaci\'{o}n} under the project MTM2010-16845 and the Junta de Extremadura Autonomous Government under the grant GR10064. 
    
    (**) This author has been supported by the spanish Ministerio de Ciencia e Innovaci\'{o}n under the project MTM2010-16660.
%\item[] \hspace*{-1cm} Proofs should be sent to: Agust\'{i}n G. Nogales, Dpto. de
%    Matem\'aticas, Universidad de Extremadura, Avda. de Elvas, s/n, 06071--Badajoz,
%    SPAIN. e-mail: nogales@unex.es
 \end{itemize}
\newpage

\section{Introduction}

Let $(\Omega,\mathcal A,P)$ be a probability space,
$X:(\Omega,\mathcal A,P)\mapsto\mathbb R^n$ be an $n$-dimensional
random variable and $Y:(\Omega,\mathcal A,P)\mapsto\mathbb R$ a
random variable with finite mean. The conditional expectation $E(Y|
X)$ is defined as a random variable on $\mathbb R^n$ such that
$\int_{X^{-1}(B)}Y\, dP=\int_BE(Y| X)dP^X$ for all Borel set
$B$ in $\mathbb R^n$, where $P^X$ denotes the probability distribution of $X$. Although the existence of the conditional
expectation is guaranteed via the Radon-Nikodym theorem, its
computation becomes, generally, a hard problem. When the joint
density $f$ of $Y$ and $X$ is known, $E(Y| X=x)$ is the mean of the
conditional distribution $P^{Y| X=x}$ of $Y$ given $X=x$, whose
density is $f(x,y)/f_X(x)$, where $f_X$ denotes the marginal
distribution of $X$. In this case the problem to compute a
conditional expectation is reduced to that to ``evaluate'' a mean,
and we have a lot of methods to do that, interpreting ``evaluation''
as ``approximation'' or ``simulation'' in a probabilistic context or ``estimation'' in
a statistical framework. When a joint density of $X$ and $Y$ is not available, or
is difficult to determine, the problem of evaluating the conditional
expectation could become an ardous problem. But this is still an
interesting problem, as $y=E(Y| X=x)$ is the regression curve of
$Y$ given $X=x$. For this reason,  many probabilistic or
statistical methods have been given to deal with, including
Monte Carlo methods or nonparametric function estimation, for instance.

Although our approach is different, the closest reference to our purposes is
Lindqvist and Taraldsen (2005), where the authors review and complement
 a general approach in a statistical context to Monte Carlo computations of conditional expectations
given  a sufficient statistic. See also the references therein.
In this paper, we describe a Monte Carlo method, inspired on a Besicovitch
theorem on differentiation of measures, to evaluate such a
conditional expectation in a probabilistic setting. Nevertheless, the method
can also be used in a statistical framework to approximate the conditional expectation given
a sufficient statistic, for instance. In fact, the method is applied in the last section
of the paper to evaluate the minimum risk equivariant estimator of
the location parameter of a general half-normal distribution. This
estimator is described in terms of a conditional expectation for known values of the location
and scale parameters that we have had to estimate by simulation. We also include in the last
section of the paper an explicit expression of the MRE estimator of
the scale parameter, which, to our knowledge, it has not been done
before. The behavior of these estimators is compared by simulation
with the behavior of maximum likelihood and unbiased estimators.

For the sake of completeness, we also give MRE estimators of the
location and scale parameters when the other is supposed to be
known, although this problem is less interesting from a point of view
of applications.

\section{A method to approximate conditional expectations}

Let us recall briefly a theorem of Besicovitch on differentiation of
measures (see, for instance, Corollary 2.14 of Mattila (1995)):

\begin{mytheo}[Besicovitch (1945, 1946)]
Let $\lambda$ be a Radon measure on $\mathbb R^n$, and $f:\mathbb
R^n\mapsto\mathbb R$ a locally $\lambda$-integrable function. Then
$$\lim_{r\downarrow 0}\frac1{\lambda(B_r(x))}\int_{B_r(x)}f\,d\lambda=f(x)
$$
for $\lambda$-almost all $x\in\mathbb R^n$, where $B_r(x)$ denotes
the ball of center $x$ and radius $r>0$ for the norm
$\|\cdot\|_\infty$ on $\mathbb R^n$.
       \end{mytheo}

Let now $(\Omega,\mathcal A,P)$ be a probability space,
$U:(\Omega,\mathcal A,P)\mapsto\mathbb R^n$ be an $n$-dimensional
random variable and $f:(\Omega,\mathcal A,P)\mapsto\mathbb R$ be a
real random variable with finite mean. Then, for $P^U$-almost every
$u\in\mathbb R^n$,
$$\lim_{\epsilon\downarrow0}\frac1{P^U(B_\epsilon(u))}\int_{U^{-1}(B_\epsilon(u))}f(\omega)\,dP(\omega)=
\lim_{\epsilon\downarrow0}\frac1{P^U(B_\epsilon(u))}\int_{B_\epsilon(u)}E(f|U=u')\,dP^U(u')=E(f|U=u)
$$

By the Strong Law of Large Numbers, for %($P^\infty$-)
almost every
sequence $(\omega_i)$ in $\Omega$, we have
\begin{gather*}
P^U(B_\epsilon(u))=\lim_k\frac1k\sum_{i=1}^kI_{B_\epsilon(u)}(U(\omega_i))\\ \mbox{ and }\\
\int_{B_\epsilon(u)}E(f|U=u')\,dP^U(u')=\lim_k\frac1k\sum_{i=1}^kI_{B_\epsilon(u)}(U(\omega_i))f(\omega_i)
\end{gather*}
Hence, we have proved the following result:

\begin{mytheo}
Let $(\Omega,\mathcal A,P)$ be a probability space,
$U:(\Omega,\mathcal A,P)\mapsto\mathbb R^n$ be an $n$-dimensional
random variable and $f:(\Omega,\mathcal A,P)\mapsto\mathbb R$ be a
real random variable with finite mean. Then, for $P^U$-almost every
$u\in\mathbb R^n$ and almost every
sequence $(\omega_i)$ in $\Omega$, we have
$$E(f|U=u)=\lim_{\epsilon\downarrow0}\lim_k\frac{\sum_{i=1}^kI_{B_\epsilon(u)}(U(\omega_i))f(\omega_i)}{\sum_{i=1}^kI_{B_\epsilon(u)}(U(\omega_i))}
$$
       \end{mytheo}

This theorem yields a way to approximate the conditional expectation
of $f$ given $U$. Let us give a simple example to illustrate the method.

\begin{myexa}\rm Let $(X,Y)$ be a bidimensional random variable normally distributed with null mean and covariance matrix
$$\left(\begin{array}{cc}
1 & 1/2\\
1/2 & 1
\end{array}\right)
$$
In this case, we don't need any approximation of the conditional expectation of $Y$ given $X=x$ because it is $x/2$. Notice that, in this simple example, the conditional distribution of $Y$ given $X=x$ is $N(\frac12 x,\frac12\sqrt{3})$. Nevertheless, if we want to apply the suggested method to calculate $E(Y|X=1)$, given a small $\epsilon>0$ small, we may choose a sample $(x_i,y_i)_{1\le i\le k}$ of the joint distribution of $X$ and $Y$ and approximate $E(Y|X=1)$ by
$$\frac{\sum_{i=1}^kI_{[1-\epsilon,1+\epsilon]}(x_i)\cdot y_i}{\sum_{i=1}^kI_{[1-\epsilon,1+\epsilon]}(x_i)}
$$
Taken $\epsilon=0.1$ and samples of the joint distribution of $X$
and $Y$ with sample sizes $k$ enough to obtain
$m=\sum_{i=1}^kI_{[1-\epsilon,1+\epsilon]}(x_i)=10, 20, 30, 50,
100$, and using the statistical software R, we have obtained the
following approximations of $E(Y|X=1)$ and box-plots (dotted red
line represents the mean) after 100 simulations:

%\begin{center}\begin{tabular}{|c|c|c|c|c|c|}\hline
%$m$ & 10 & 20 & 30 & 50 & 100\\\hline
%$E(Y|X=1)$ & 0.501 & 0.521 & 0.504 & 0.521 & 0.511\\\hline
%\end{tabular}\vspace{2ex}\\
%\scalebox{1}{\includegraphics[width=200pt,height=200pt]{Grafexy2.pdf}}
%\end{center}

\begin{center}\begin{tabular}{|c|c|c|c|c|c|}\hline
$m$ & 10 & 20 & 30 & 50 & 100\\\hline $E(Y|X=1)$ & 0.5007 & 0.5211 &
0.5037 & 0.5211 & 0.5114\\\hline
\end{tabular}\vspace{1ex}\\
Table 1. Approximation  of $E(Y|X=1)$ ($\epsilon=0.1$, $=10, 20, 30,
50, 100$, 100 simulations).\vspace{2ex}\\
\scalebox{.65}{\includegraphics[width=200pt,height=200pt]{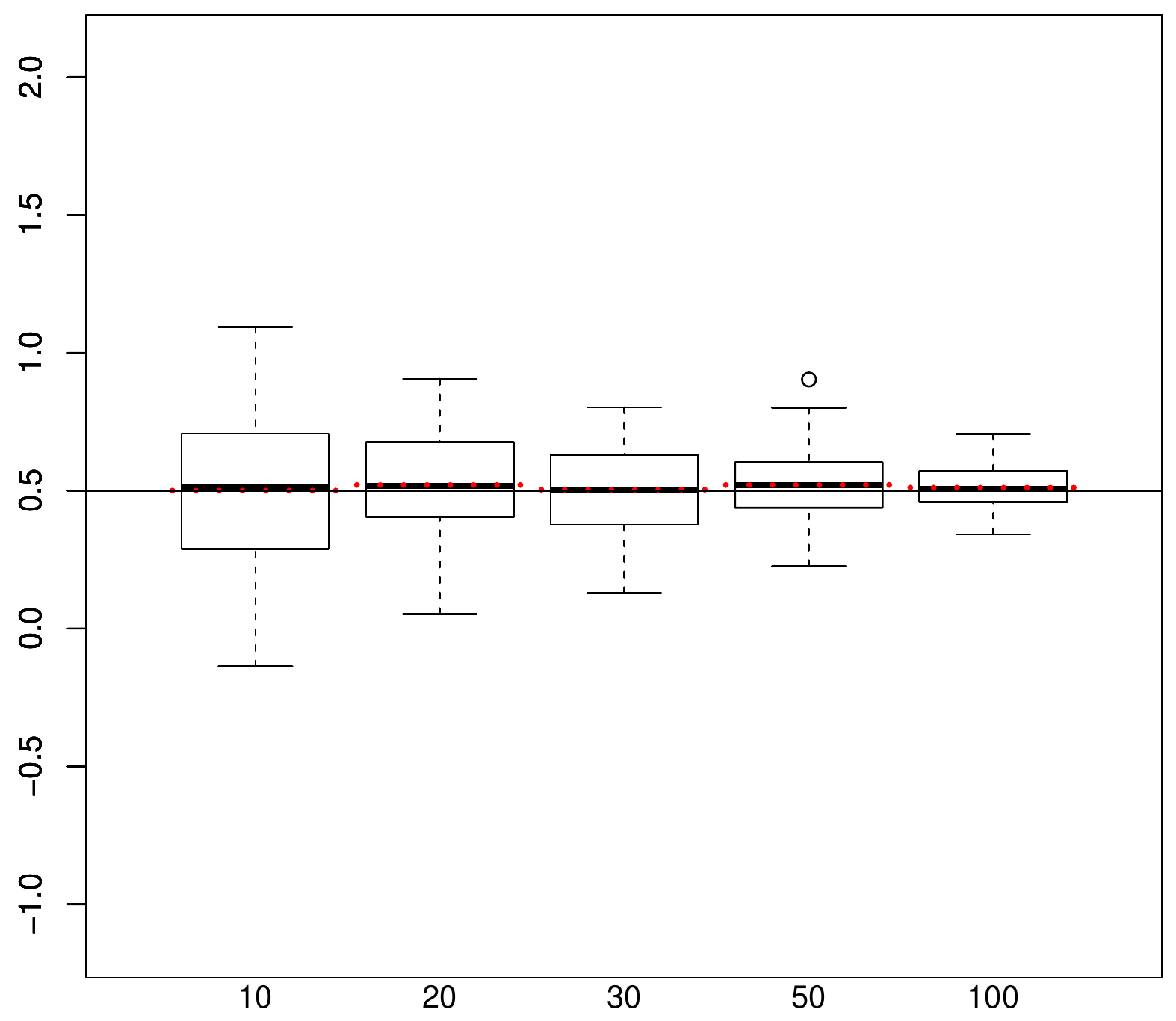}}\vspace{1ex}\\
Figure 1. Box plots of the approximations  of $E(Y|X=1)$\\
($\epsilon=0.1$, $=10, 20, 30, 50, 100$, 100
simulations).\end{center}

%A method to approximate such a conditional mean, useful when its calculus is not easy, is the following:  choose an $n$-sized sample $x_1,\dots,x_n$ of the uniform distribution on $[1-\epsilon,1+\epsilon]$ (that, for $\epsilon>0$ small, don't should differ of a sample of the distribution of $X$ on this interval) and then, for each $1\le i\le n$, choose a value $y_i$ at
%random of the distribution $N(\frac12 x_i,\frac12\sqrt{3})$, so that the approximation of $E(Y|X=1)$ becomes
%$$\frac1n\sum_{i=1}^n y_i.
%$$

A similar simulation study has been performed to approximate the conditional expectation $E(V|U=0.5)$, where $V=\sin(X\cdot Y)$ and $U=\cos(X^2+Y^2)$; the obtained results are:

%\begin{center}\begin{tabular}{|c|c|c|c|c|c|}\hline
%$m$ & 10 & 20 & 30 & 50 & 100\\\hline
%$E(V|U=0.5)$ & 0.13 & 0.12 & 0.13 & 0.13  & 0.13\\\hline
%\end{tabular}\\
%\scalebox{1}{\includegraphics[width=200pt,height=200pt]{Grafsencos.pdf}}
%\end{center}

\begin{center}\begin{tabular}{|c|c|c|c|c|c|}\hline
$m$ & 10 & 20 & 30 & 50 & 100\\\hline $E(V|U=0.5)$ & 0.1235 & 0.1058
& 0.1309 & 0.1341  & 0.1281\\\hline
\end{tabular}\vspace{1ex}\\
Table 2. Approximation  of $E(V|U=0.5)$ ($\epsilon=0.1$, $=10, 20,
30, 50, 100$, 100 simulations).\vspace{2ex}\\

\scalebox{.65}{\includegraphics[width=200pt,height=200pt]{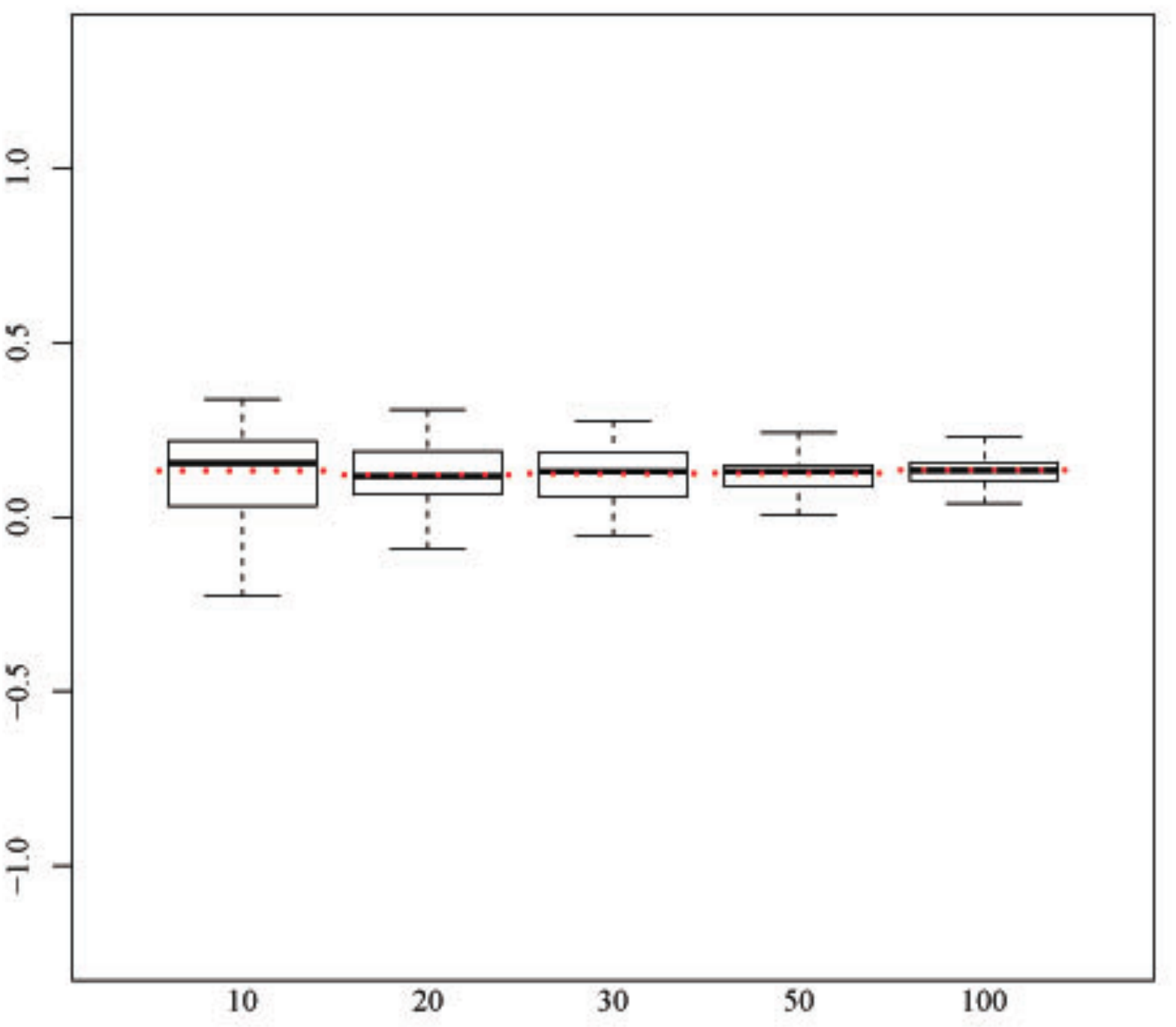}}\vspace{1ex}\\
Figure 2. Box plots of the approximations  of $E(V|U=0.5)$\\
($\epsilon=0.1$, $=10, 20, 30, 50, 100$, 100 simulations)
\end{center}
\hfill $\Box$
\end{myexa}
Notice that in the example $k$ should be an enough great number to
secure a good size $m$ of non-null terms in the denominator of this
expression. Besides, the smaller $\epsilon$, greater has to be $k$.
This may become a problem when this method is applied, especially
when $X$ is a random vector of high dimension. Any additional
information about the distribution of $X$ may be useful in some way
to circumvent this problem, as indeed occur when determining the
minimum risk equivariant estimator (MRE) of the location parameter
$\xi$ of the general half-normal distribution in the next section.

\section{Application to equivariant estimation of the location parameter of the general half-normal distribution}

  Let $Z$ be a real random variable (r.r.v.) with distribution $N(0,1)$. The distribution of the
r.r.v. $X:=|Z|$ is the so-called half-normal distribution. It will
be denoted $HN(0,1)$ and its density function is
$$f_X(x)=\sqrt{\frac{2}{\pi}}\exp\left\{-\frac12x^2\right\}I_{[0,+\infty[}(x).
$$

A general half-normal distribution $HN(\xi,\eta)$ is obtained from
$HN(0,1)$ by a location-scale transformation: $HN(\xi,\eta)$ is the
distribution of $Y=\xi+\eta X$.\par

The classical paper Daniel (1959) introduces half-normal plots and
the half-normal distribution. The half-normal distribution is a
special case of the folded normal and truncated normal distribution
(see Johnson et al. (1994)).  Bland et al. (1999) and  Bland (2005)
propose a so-called half-normal method to deal with relationships
between measurement error and magnitude, with applications in
medicine.
%Cooray et al. (2008) considers another generalization of
%the half-normal distribution and gives some applications to lifetime
%data.
Pewsey (2002) uses the maximum likelihood principle to estimate the
parameters, and contains a brief survey on the general half-normal
distribution, its relations with other well-known distributions and
its usefulness in the analysis of highly skew data;  Pewsey (2004)
proposes bias-corrected estimators of the estimators quoted before.
%Farsipour et al. (2006) and Wiper et al. (2008) consider inference
%problems for these parameters in a Bayesian framework.
 Nogales et
al. (2011) deals with the problem of unbiased estimation in the
general half-normal distribution. This paper is mainly devoted to
the problem of equivariant estimation of the location and scale
parameters, $\xi$ and $\eta$, but first we do a brief review on the
results about unbiased and maximum likelihood estimation appearing
in the literature.

%But, first, let us recall we do a brief review on the results
%about unbiased and maximum likelihood estimation appearing in the
%literature.
%
%
%
%\section{Unbiased and maximum likelihood estimation of $\xi$ and $\eta$}

The density function of $HN(\xi,\eta)$ is

$$f_Y(y)=\frac1{\eta} f_X\left(\frac{y-\xi}{\eta}\right)=\frac1{\eta} \sqrt{\frac2{\pi}}
\exp\left\{-\frac12 \left(\frac{y-\xi}{\eta}\right)^2
\right\}I_{[\xi,+\infty[}(y).
$$
It is readily shown that
$$E(Y)=\xi+\eta\sqrt{\frac{2}{\pi}}\qquad\text{and}\qquad \mbox{Var}(Y)=\frac{\pi-2}{\pi}\eta^2.
$$

Let $Y_1,\dots,Y_n$ be a sample of size $n$ from a general half-normal distribution with unknown parameters, $\xi$ and $\eta$.
$Y_{1:n}$ denotes the minimum of $Y_1,\dots,Y_n$. From the factorization criterion, we obtain that $(\sum_{i=1}^nY_i^2,\sum_{i=1}^nY_i,Y_{1:n})$ is a sufficient statistic. Indeed, it is minimal sufficient, although not complete.

We write $Y_i=\xi+\eta X_i$, where $X_i=|Z_i|$, $1\le i\le n$, $Z_1,\dots,Z_n$ being a sample of the standard normal distribution
$N(0,1)$. Throughout this paper, we also write $$c_n:=E(X_{1:n})$$

For $n\ge 2$, it is readily shown that $0<c_n<\sqrt{\frac{2}{\pi}}$.
In fact, the next lemma  (Nogales el al. (2011)) yields an alternative expression and a
refined bound for $c_n$. We write $\Phi$ for the standard normal
cumulative distribution function.

\begin{mylem} \rm \label{l1}

{\rm (i)} $c_n=\int_0^{\infty}(2-2\Phi(t))^n\,dt$.

{\rm (ii)} For $n\ge 1$, $c_n\le\frac1n\sqrt{\frac{\pi}{2}}\le
\Phi^{-1}\left(\frac12+\frac1{2n}\right)$.\end{mylem}

Notice also that $Y_{1:n}=\min_iY_i=\xi+\eta X_{1:n}$ and $E(Y_{1:n})=\xi+\eta c_n$.

The next proposition (Nogales el al. (2011)) yields unbiased
estimators of the location and scale parameters, $\xi$ and $\eta$.
Both estimators are $L$-statistics and function of the minimal
sufficient statistic cited.

\begin{myprop}  %\rm

{\rm (i)} $\widetilde\xi:=\frac{\sqrt{\frac{2}{\pi}}Y_{1:n}-
c_n{\bar Y}}{\sqrt{\frac{2}{\pi}}-c_n}$ is an unbiased estimator of
the location parameter~$\xi$.

{\rm (ii)} $\widetilde\eta:=\frac{{\bar
Y}-Y_{1:n}}{\sqrt{\frac{2}{\pi}}-c_n}$ is an unbiased estimator of
the scale parameter $\eta$ whose distribution does not depend on
$\xi$.

\end{myprop}

\begin{myrema}\rm  We also have that the sample mean ${\bar Y}$ is an unbiased estimator of the mean $\xi+\eta\sqrt{\frac{2}{\pi}}$. Moreover, an unbiased estimator of $\eta^2$ is
$$\frac{\pi}{\pi-2}\,S^2,
$$
where $S^2:=\frac1{n-1}\sum_{i=1}^n(Y_i-{\bar Y})^2$ is the sample
variance; notice that its distribution does not depend on $\xi$.
${\bar Y}$ and $S^2$ also are functions of the sufficient statistic
given above. The reader is referred to Nogales et al. (2011) for
these and other results about unbiased estimation of the parameters
of the general half-normal distribution. $\Box$ \end{myrema}

\begin{myrema}\rm  Pewsey (2002) provides maximum likelihood estimates for each of the parameters $\xi$ and $\eta$:
$$\widehat{\xi}:=Y_{1:n},\quad \widehat{\eta}:=\left(\frac1n\sum_{i=1}^n(Y_i-Y_{1:n})^2\right)^{1/2}
$$
A large sample based bias-correction is used in Pewsey (2004) to
improve the performance of the maximum likelihood estimators
$\widehat\xi$ and $\widehat\eta$. $\Box$ \end{myrema}

In this section we consider the problem of determining the minimum
risk equivariant estimator of the position
parameter $\xi$ of the general half-normal distribution
$HN(\xi,\eta)$ when the scale parameter $\eta$ is unknown. We cannot
provide an explicit expression for this estimator, since it is
described in terms of two conditional expectations that had to be
estimated by simulation. To achieve this goal, an R program has been
developed based on the method of the previous section.

For the sake of completeness, we also give MRE estimators of the
scale parameter, and of one of the parameters when the other is
supposed to be known since, as far as we know, they have not been
yet reported in the literature. The results are a consequence of the
classical equivariant estimation theory, as it appears, for
instance, in Lehmann (1983).

%\section{Equivariant estimation of scale parameter when both parameters are unknown}

To estimate the location parameter $\xi$ when the scale parameter
$\eta$ is unknown, we have the next result (a direct consequence of
Lehmann (1986, p. 182)).

\begin{myprop} %\rm
When the loss function $W_2(x;\xi,\eta)=\eta^{-2}(x-\xi)^2$ is considered,
the MRE estimator $\overset{\circ}{\xi}$ of $\xi$ is
$$\overset{\circ}{\xi}=T^*_0-(\rho\circ U)\cdot T^*_1
$$
where
\begin{gather*}
T^*_0=\bar Y,\quad T^*_1=\frac1n\sum_{i=1}^n|Y_i-\bar Y|\\
U=\left(\frac{Y_1-Y_n}{Y_{n-1}-Y_n},\dots,\frac{Y_{n-2}-Y_n}{Y_{n-1}-Y_n},\frac{Y_{n-1}-Y_n}{|Y_{n-1}-Y_n|}\right),\\
%Z_i=\frac{X_i-X_n}{X_{n-1}-X_n} \mbox{ for } i=1,\dots,n-2,\quad Z_{n-1}=\frac{X_{n-1}-X_n}{|X_{n-1}-X_n|}\\
%Y_i= \mbox{ for } i=1,\dots,n-1, \mbox{ and}\\
\rho=\frac{E_{\xi=0,\eta=1}[T^*_0\cdot
T^*_1|U]}{E_{\xi=0,\eta=1}[{T^*_1}^2|U]}
\end{gather*}
\end{myprop}

\begin{myrema}\rm
$T^*_0$ can be replaced by any other equivariant estimator of $\xi$,
and $T^*_1$ can be replaced by any positive estimator of $\eta$
satisfying $T^*_1(a+by_1,\dots,a+by_1)=b T^*_1(y_1,\dots,y_1)$ for
every $a\in\mathbb R$, $b>0$. $\Box$ \end{myrema}

\begin{myrema}\rm
A simulation was realized to visualize the behavior of the minimum risk equivariant estimator
$\overset{\circ}{\xi}$. For this simulation we did 100 simulations
with sample sizes $n=10,20,30,50,100$ of a half-normal distribution $HN(10,4)$
obtaining the next results:
%\begin{center}
%\begin{tabular}{|c|c|c|c|c|c|}
%\hline
%%  &  \multicolumn{2}{c|}{$\tilde{\xi}$ }  &  \multicolumn{2}{c|}{$\hat{\xi}$ }  &  \multicolumn{2}{c|}{$\overset{\circ}{\xi}$ }\\ \hline
% $n$ &  10 &  20 &  30 & 50 & 100 \\ \hline
%Mean   & 9.87 & 9.60 & 9.42 & 9.54 & 9.64 \\ \hline
%  MSE & 1.07 & 0.93 & 1.62 & 0.90 & 0.41 \\ \hline
%\end{tabular}\\
%Table 1. Simulated mean and MSE of the estimators  $\overset{\circ}{\xi}$.\vspace{2ex}\\
%\scalebox{1.2}{\includegraphics[width=200pt,height=200pt]{Grafxicirc.pdf}}
%\end{center}

\begin{center}
\begin{tabular}{|c|c|c|c|c|c|}
\hline
%  &  \multicolumn{2}{c|}{$\tilde{\xi}$ }  &  \multicolumn{2}{c|}{$\hat{\xi}$ }  &  \multicolumn{2}{c|}{$\overset{\circ}{\xi}$ }\\ \hline
 $n$ &  10 &  20 &  30 & 50 & 100 \\ \hline
Mean   & 9.8710 & 9.6038 & 9.4242 & 9.5351 & 9.6429 \\ \hline
  MSE & 1.0684 & 0.9301 & 1.6223 & 0.9041 & 0.4050 \\ \hline
\end{tabular}\vspace{1ex}\\
Table 3. Simulated mean and MSE of the estimators  $\overset{\circ}{\xi}$.\vspace{2ex}\\
\scalebox{.7}{\includegraphics[width=200pt,height=200pt]{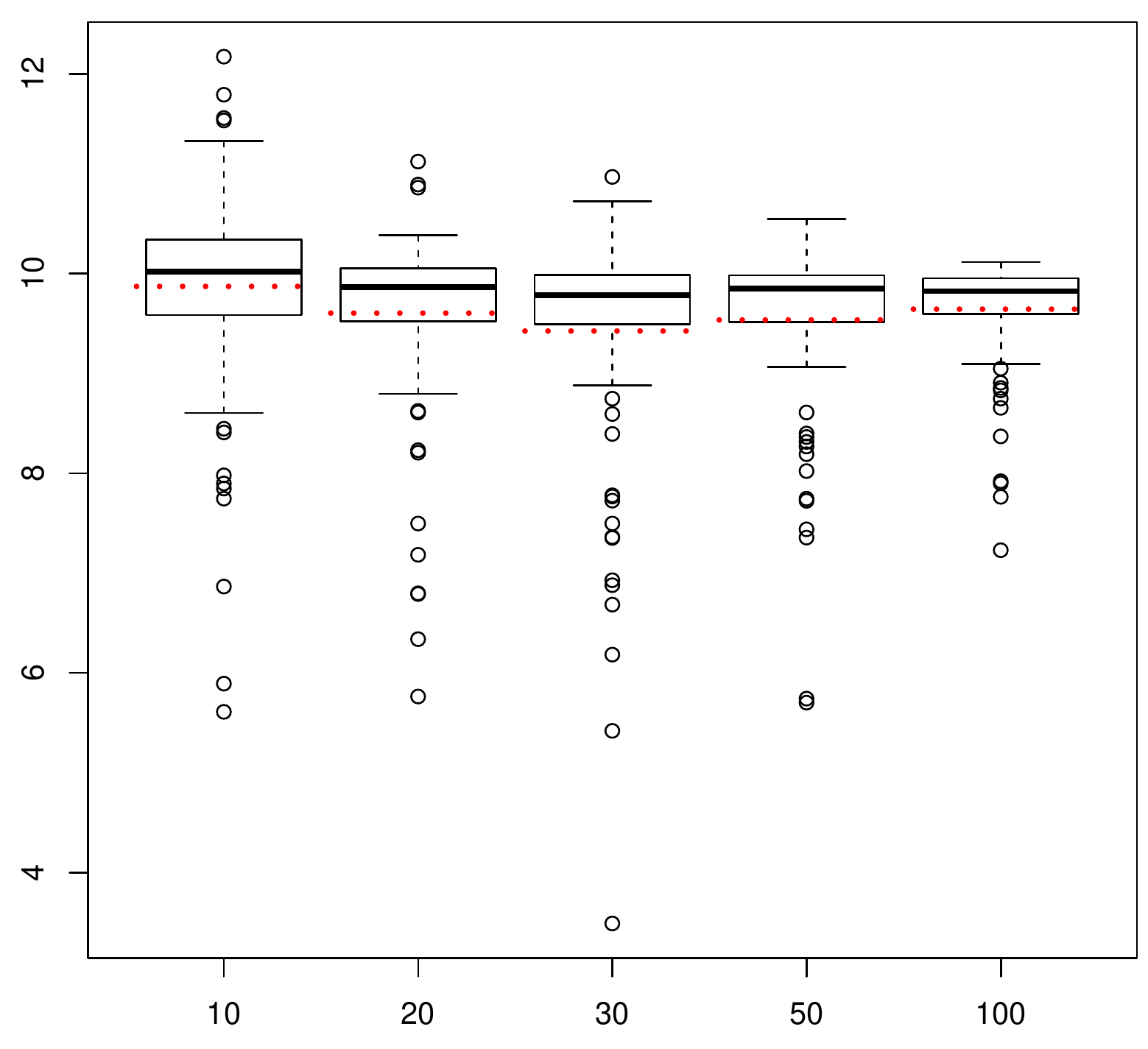}}\vspace{1ex}\\
Figure 3. Box plots of the simulations.
\end{center}

To compare the behavior of the unbiased estimator $\tilde{\xi}$, the maximum likelihood estimator
$\hat{\xi}$ and the minimum risk equivariant estimator
$\overset{\circ}{\xi}$, we did 100 simulations
with sample sizes $n=100$ of a half-normal distribution $HN(10,4)$
obtaining the next results:

%\begin{center}
%\begin{tabular}{|c|c|c|c|c|c|c|}
%\hline
%  &  \multicolumn{2}{c|}{$\tilde{\xi}$ }  &  \multicolumn{2}{c|}{$\hat{\xi}$ }  &  \multicolumn{2}{c|}{$\overset{\circ}{\xi}$ }\\ \hline
% $n$ &  Mean &  MSE &  Mean & MSE & Mean  & MSE \\
% \hline
%  100 & 9.996 & 0.001 & 10.046 & 0.004 & 9.643 & 0.405 \\ \hline
%\end{tabular}\\
%Table 1. Simulated mean and MSE of the estimators $\tilde{\xi}$, $\hat{\xi}$ and $\overset{\circ}{\xi}$.\vspace{2ex}\\
%\scalebox{1.2}{\includegraphics[width=200pt,height=200pt]{Grafxihattildecirc.pdf}}
%\end{center}

\begin{center}
\begin{tabular}{|c|c|c|c|c|c|c|}
\hline
  &  \multicolumn{2}{c|}{$\tilde{\xi}$ }  &  \multicolumn{2}{c|}{$\hat{\xi}$ }  &  \multicolumn{2}{c|}{$\overset{\circ}{\xi}$ }\\ \hline
 $n$ &  Mean &  MSE &  Mean & MSE & Mean  & MSE \\
 \hline
  100 & 9.9964 & 0.0018 & 10.0457 & 0.0039 & 9.6429 & 0.4050 \\ \hline
\end{tabular}\vspace{1ex}\\
Table 4. Simulated mean and MSE of the estimators $\tilde{\xi}$, $\hat{\xi}$ and $\overset{\circ}{\xi}$.\vspace{2ex}\\
\scalebox{.7}{\includegraphics[width=200pt,height=200pt]{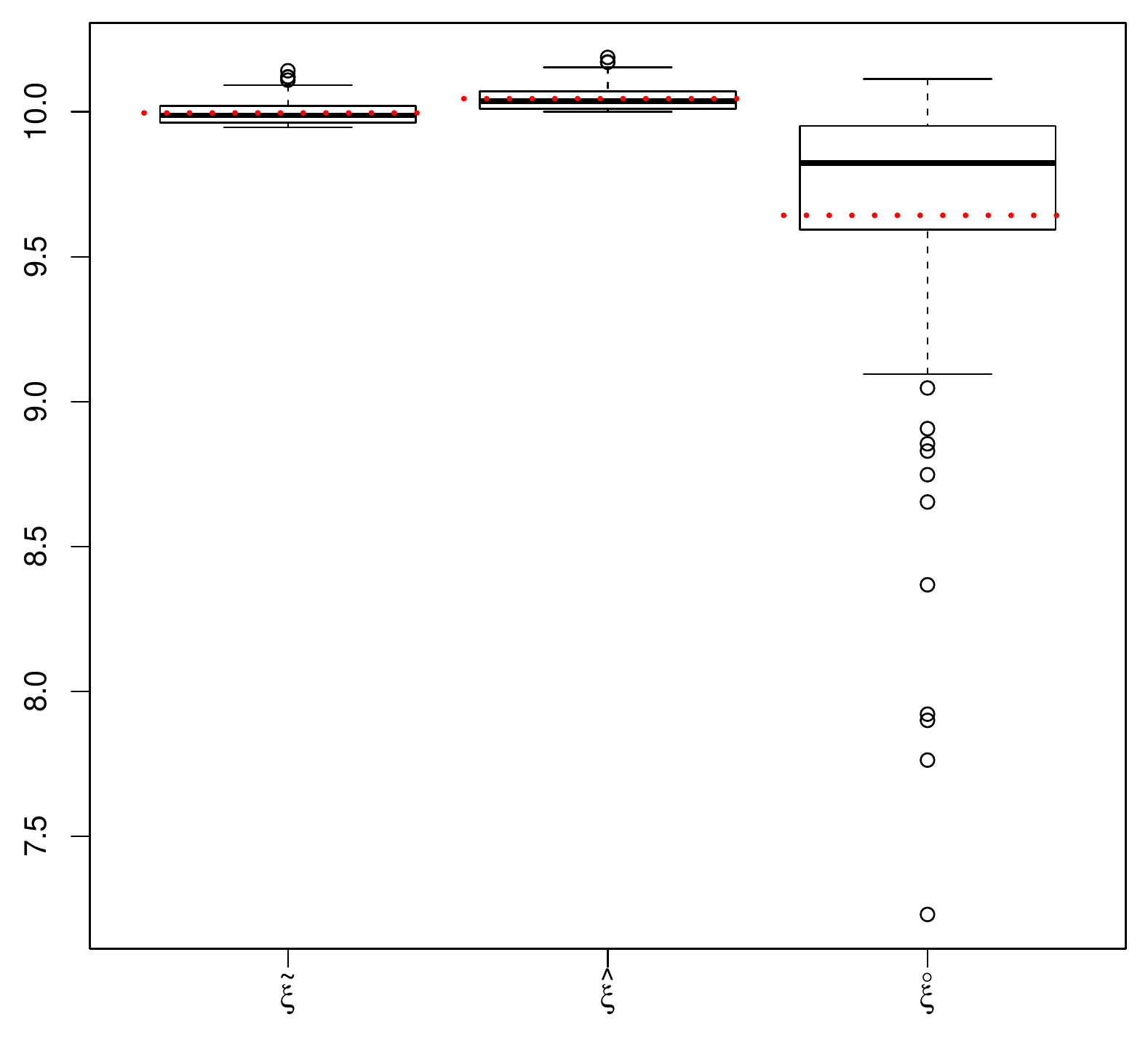}}\vspace{1ex}\\
Figure 4. Box plots of the simulations.
\end{center}

We can see the biased character of the maximum likelihood estimator
$\hat{\xi}$ and the minimum risk equivariant estimator
$\overset{\circ}{\xi}$. Obviously, as it can be expected, the behavior of this approximation of the MRE estimator is worse than those of the unbiased estimator $\tilde{\xi}$ or the maximum likelihood estimator $\hat{\xi}$. However, this method provides a way to proceed when other estimation methods are not available.

%in the following graph:
%
%\begin{figure}[ht!]
%   \centering
%   %%----primera subfigura----
%   %\subfloat[]{
%       %\label{fig:museo:a}         %% Etiqueta para la primera subfigura
%        \includegraphics[width=60mm, angle=90]{boxplot_medias}\\[-25pt]
%   %\hspace{0.1\linewidth}
%   %%----segunda subfigura----
%   %\subfloat[]{
%       %\label{fig:museo:b}         %% Etiqueta para la segunda subfigura
%        %\includegraphics[width=60mm, angle=90]{boxplot_ES}\
%   \caption{Estimation of the location parameter $\xi$ of a $HN(\xi,\eta)$ distribution}
%   \label{fig:museo}                %% Etiqueta para la figura entera
%\end{figure}

This simulation was performed with the statistical program R.

Let us summarize the idea used in this estimation: for a sample
$y=(y_1,\dots,y_n)$, $n=10, 20, 30, 50, 100$, of the distribution $HN(10,4)$, we have
$$\rho(U(y))=\lim_{\epsilon\to 0}\frac{N_\epsilon}{D_\epsilon}
$$
where
\begin{gather*}
N_\epsilon=\int_{A_\epsilon(y)}f(y')dy',\quad D_\epsilon=\int_{A_\epsilon(y)}g(y')dy'\\
f(y')=T^*_0(y')\cdot T^*_1(y')\cdot \exp\left\{-\frac12\|y'\|_2^2\right\},\quad g(y')=T^*_1(y')^2\cdot \exp\left\{-\frac12\|y'\|_2^2\right\}\\
A_\epsilon(y)=\{y'\in[0,10]^n\colon \max_{1\le i\le
n}|U_i(y')-U_i(y)|\le\epsilon\}
\end{gather*}
Now, take a sample $S$ of $A_\epsilon(y)$ and approximate $N_\epsilon$
and $D_\epsilon$ by $\frac1{\mbox{card\,}(S)}\sum_{y'\in S}f(y')$
and $\frac1{\mbox{card\,}(S)}\sum_{y'\in S}g(y')$, resp. So,
$\rho(U(y))$ can be estimated by
$$C(y):=\frac{\sum_{y'\in S}f(y')}{\sum_{y'\in S}g(y')}
$$
and $\overset{\circ}{\xi}(y)$ is approximated by $D(y):=T^*_0(y)-C(y)\cdot T^*_1(y)$.

To approximate $C(y)$, a first idea would be to divide the interval $[0,10]$ in multiple subintervals of small length $\epsilon>0$ and consider the grid in the interval $[0,10]^n$ formed by the $n$-power set of the ends of these subintervals (we have restricted ourselves to the interval [0,10] because we have considered virtually nil the functions $ f (y) $ and $ g (y) $ when one of the coordinates of the vector $y$ is greater than 10). Sample $S$ would be formed by the grid nodes that are in $A_\epsilon$. The main problem with this approach is that the size $m$ of the sample $S$ is very small (it becomes smaller when the greater is the dimension $n$). To secure a sample size $m$ enough for $S$ (given $n$, we take $m=100\cdot n$), we have used the following algorithm, that benefits from the invariance of $U$ under scale and location transformations:
\begin{itemize}
\item Given a sample $y=(y_1,\dots,y_n)$ of the distribution $HN(10,4)$, take $w_{n-1},w_{n}$ at random in
$[0,10]$ such that $w_{n-1}-w_{n}$ has the same sign than $y_{n-1}-y_{n}$.

\item For $1\le i\le n-2$, let $a_i:=\frac{y_{1}-y_{n}}{y_{n-1}-y_{n}}$ and take $0<\epsilon
<\min\{0.1, \min_{1\le i\le n-2}|a_i|\}$.

\item For $1\le i\le n-2$ take $w_i$ at random on the interval determined by $w_n+(w_{n-1}-w_{n})(a_i-\epsilon)$ and
$w_n+(w_{n-1}-w_{n})(a_i+\epsilon)$.

\item The process is repeated until $100\cdot n$ vectors $w^{(j)}=(w^{(j)}_1,\dots,w^{(j)}_n)$,
$1\le j\le 100\cdot n$ are obtained.

\item If $w^{(j_0)}_{i_0}<0$ for some $i_0,j_0$, we replace $w^{(j)}_i$, $1\le i\le n$,
$1\le j\le 100\cdot n$, by $v+w^{(j)}_i$, where $v$ is choosen at random between
$-\min_{1\le i\le n, 1\le j\le 100\cdot n}w^{(j)}_i$ and
$1-\min_{1\le i\le n, 1\le j\le 100\cdot n}w^{(j)}_i$.

\item Each new $w^{(j)}$ is divided by $\max_{1\le i\le n}w^{(j)}_i$ and multiplied by a random number
choosen in $[0,10]$.

\item Take $S=\{w^{(j)}\colon 1\le j\le 100\cdot n\}$.

\end{itemize}

Finally,  we choose $k:=100$ samples $y^{(i)}$ of size $n$ of the distribution $HN(10,4)$
and estimate the mean of $\overset{\circ}{\xi}$ by
$$\frac1k\sum_{i=1}^kD(y^{(i)})
$$
and the mean squared error  $\overset{\circ}{\xi}$ by
$$\frac1k\sum_{i=1}^k(D(y^{(i)})-10)^2.
$$

$\Box$
\end{myrema}

\begin{myrema}\rm
When the scale parameter $\eta$ is supposed known (say
$\eta=\eta_0$), the joint density of $Y_1,\dots,Y_n$ is
$$f_\xi(y_1,\dots,y_n)=\frac1{\eta_0^n} \sqrt{\frac2{\pi}}^{\;n}
\exp\left\{-\frac1{2\eta_0^2} \sum_{i=1}^n(y_i-\xi)^2
\right\}I_{[\xi,+\infty[}(y_{1:n}),
$$
where $y_{1:n}:=\min\{y_1,\dots,y_n\}$. This family remains
invariant under translations of the form
$g_a(y_1,\dots,y_n)=(y_1-a,\dots,y_n-a)$.

The equivariant estimator of minimum mean squared error of the
location parameter $\xi$ is
$$T_1={\bar Y}-\frac{\eta_0}{\sqrt{2\pi n}}\frac{\exp\left\{-\frac{n}{2\eta_0^2}\left(Y_{1:n}-{\bar Y}\right)^2\right\}}
{\Phi\left[\frac{\sqrt{n}}{\eta_0}\left(Y_{1:n}-{\bar
Y}\right)\right]}.
$$

In fact, for the loss function $W'_2(\xi,x)=(x-\xi)^2$, the MRE
estimator of the location parameter $\xi$ is the Pitman estimator
$$T_1(y_1,\dots,y_n)=\frac{\int_{-\infty}^{+\infty}uf_0(y_1-u,...,y_n-u)du}
{\int_{-\infty}^{+\infty}f_0(y_1-u,...,y_n-u)du}
$$
For $y\in\mathbb {R}^n$, we write $\bar y$ for the mean of
$y_1,\dots, y_n$. After some algebraic manipulations, we obtain:

\begin{gather*}
\int_{-\infty}^{+\infty}uf_0(y_1-u,...,y_n-u)du=\\
\begin{split}\left(\frac{\sqrt{2}} {\eta_0\sqrt{\pi}}\right)^n
&\exp\left\{-\frac{1}{2\eta_0^2}\left(\sum_{i=1}^n y_i^2-n{\bar
y}^2\right)\right\}\frac{\eta_0}{\sqrt{n}}\\
&\cdot\left[-\frac{\eta_0}{\sqrt{n}}\exp\left\{-\frac{n}{2\eta_0^2}(y_{1:n}
-{\bar y})^2\right\}+{\bar y}\sqrt{2\pi}\,
\Phi\left(\frac{\sqrt{n}}{\eta_0} (y_{1:n}-{\bar y})\right)\right]
\end{split}\end{gather*}
and
\begin{gather*}
\int_{-\infty}^{+\infty}f_0(y_1-u,...,y_n-u)du=\\
\left(\frac{\sqrt{2}}{\eta_0\sqrt{\pi}}\right)^n
\exp\left\{-\frac{1}{2\eta_0^2}\left(\sum_{i=1}^n y_i^2-n{\bar
y}^2\right)\right\}\frac{\eta_0}{\sqrt{n}} \sqrt{2\pi}\,
\Phi\left[\frac{\sqrt{n}}{\eta_0} (y_{1:n}-{\bar y})\right]
\end{gather*}
and the statement follows easily from these expressions. $\Box$
\end{myrema}

Unlike what happens with the location parameter $\xi$, for the scale
parameter $\eta$ an explicit expression for the MRE estimator is
obtained.

We consider the scale-location family of densities
$$f_{(\xi,\eta)}(y_1,...,y_n)=\frac{1}{\eta^n}f\left(\frac{y_1-\xi}{\eta},...,
\frac{y_n-\xi}{\eta}\right),$$ where
$$f(y_1,...,y_n)=
\left(\frac{2}{\pi}\right)^{\frac{n}{2}}
\exp\left\{-\frac{1}{2}\sum_{i=1}^n y_i^2\right\}\cdot
I_{[0,+\infty[}(y_{1:n}).$$ This family remains invariant under
transformations of the form
$g_{a,b}(y_1,...,y_n)=(a+by_1,...,a+by_n)$, $a\in\mathbb{R}$, $b>0$.

\begin{myprop} %\rm
When using the loss function $W_1(x;\xi,\eta)=\eta^{-2}(x-\eta)^2$,
the MRE estimator $\overset{\circ}{\eta}$ of $\eta$ is
$$\overset{\circ}{\eta}(y)=\sqrt{\frac{n-1}{2}}\cdot\frac{\Gamma\left(\frac{n+1}{2}\right)}{\Gamma\left(\frac{n+2}{2}\right)}
\cdot\frac{t(n+1)\left(\left[\sqrt{\frac{n(n+1)}{n-1}}\frac{\bar
y-y_{1:n}}{S(y)},\infty\right[\right)}{t(n+2)\left(\left[\sqrt{\frac{n(n+2)}{n-1}}\frac{\bar
y-y_{1:n}}{S(y)},\infty\right[\right)}\cdot S(y)
$$
where $t(n)$ denotes Student's $t$-distribution with $n$
degrees of freedom and $S^2$ is the sample variance.
\end{myprop}

\begin{myproo} \rm
The MRE estimator of the scale parameter $\eta$, when using the
loss function $W_1$, is
$$\overset{\circ}{\eta}(y)=\frac{\int_0^{+\infty}v^nf'(vy'_1,...,vy'_{n-1})dv}
{\int_0^{+\infty}v^{n+1}f'(vy'_1,...,vy'_{n-1})dv},$$ where $f'$ is
the joint density when $\eta=1$ of $Y'_i:=Y_i-Y_n$, $1\le i\le n-1$,
and $y'_i:=y_i-y_n$, $1\le i\le n-1$.

Notice that

\begin{gather*}
f'(y'_1,...,y'_{n-1})=\int_{-\infty}^{+\infty}f(y_1+t,...,y_n+t)dt\\
=\left(\frac{2}{\pi}\right)^{\frac{n}{2}}\exp\left\{-\frac12\sum_{i=1}^ny_i^2+\frac{n}{2}{\bar y}^2\right\}\int_{-y_{1:n}}^\infty\exp\left\{-\frac{n}{2}(t+\bar y)^2\right\}dt\\
=\frac1{\sqrt{n}}\left(\frac{2}{\pi}\right)^{\frac{n}{2}}\exp\left\{-\frac12(n-1)S^2(y)\right\}\int_{\sqrt{n}(\bar y-y_{1:n})}^\infty\exp\left\{ -\frac12u^2\right\}du
\end{gather*}

Hence, for $k\in\mathbb{N}$, applying Fubini's Theorem after a suitable change of variables in the inner integral,
\begin{gather*}
I_k(y):=\int_0^{\infty}v^kf'(vy'_1,...,vy'_{n-1})dv\\=\frac1{\sqrt{n}}\left(\frac{2}{\pi}\right)^{\frac{n}{2}}
\int_0^\infty v^k\exp\left\{-\frac12(n-1)v^2S^2(y)\right\}\int_{\sqrt{n}(\bar y-y_{1:n})}^\infty\exp\left\{ -\frac12u^2\right\}dudv\\
=\frac1{\sqrt{n}}\left(\frac{2}{\pi}\right)^{\frac{n}{2}}
\int_{\sqrt{n}(\bar y-y_{1:n})}^\infty J_k(t,y)dt\end{gather*}
where
\begin{gather*}
J_k(t,y):=\int_0^\infty v^{k+1}\exp\left\{-\frac12v^2(t^2+(n-1)S^2(y))\right\} dv=\frac{2^{k/2}\Gamma\left(\frac{k+2}{2}\right)}{(t^2+(n-1)S^2(y))^{\frac{k+2}{2}}}
\end{gather*}
where, for $t\ge \sqrt{n}(\bar y-y_{1:n})$, we have made the change of variables $w=\frac12v^2(t^2+(n-1)S^2(y))$.

So
\begin{gather*}
I_k(y)=\frac1{\sqrt{n}}\left(\frac{2}{\pi}\right)^{\frac{n}{2}}2^{k/2}\Gamma\left(\frac{k+2}{2}\right)
\int_{\sqrt{n}(\bar y-y_{1:n})}^{\infty}\frac{dt}{(t^2+(n-1)S^2(y))^{\frac{k+2}{2}}}\\
=\frac{2^{\frac{n+k}{2}}\Gamma\left(\frac{k+1}{2}\right)}{\sqrt{n}\pi^{\frac{n-1}{2}}(n-1)^{\frac{k+1}{2}}S(y)^{k+1}}\cdot t(k+1)\left(\left[\sqrt{\frac{n(k+1)}{n-1}}\frac{\bar y-y_{1:n}}{S(y)},\infty\right[\right)
\end{gather*}
 Finally
$$\overset{\circ}{\eta}(y)=\frac{I_n(y)}{I_{n+1}(y)}=\sqrt{\frac{n-1}{2}}\frac{\Gamma\left(\frac{n+1}{2}\right)}{\Gamma\left(\frac{n+2}{2}\right)}
\frac{t(n+1)\left(\left[\sqrt{\frac{n(n+1)}{n-1}}\frac{\bar
y-y_{1:n}}{S(y)},\infty\right[\right)}{t(n+2)\left(\left[\sqrt{\frac{n(n+2)}{n-1}}\frac{\bar
y-y_{1:n}}{S(y)},\infty\right[\right)}\cdot S(y)
$$
$\Box$ \end{myproo}

\begin{myrema}\rm  To compare the behavior of the unbiased estimator
$\tilde\eta$, the maximum likelihood estimator $\hat\eta$ and the
MRE estimator $\overset{\circ}{\eta}$, we have made a simulation
study for different sample sizes $(n=10,20,30)$ of a general
half-normal distribution $HN(10,4)$; from 10000 values of the
corresponding estimators we have simulated its mean and its mean
squared error (MSE). The next table contains the results:

\begin{center}\begin{tabular}{|c|c|c|c|c|c|c|}\hline
 & \multicolumn{2}{c|}{$\tilde{\eta}$} & \multicolumn{2}{c|}{$\hat{\eta}$} & \multicolumn{2}{c|}{$\overset{\circ}{\eta}$} \\\hline
 $n$ & Mean & MSE  & Mean & MSE  & Mean & MSE \\\hline
 10 & 3.9977 & 1.1024 & 3.5220 & 1.0055 & 3.5698 & 0.9832 \\\hline
 20 & 3.9940 & 0.4932 & 3.7595 & 0.4487 & 3.7942 & 0.4404 \\\hline
 30 & 3.9971 & 0.3255 & 3.8339 & 0.2909 & 3.8600 & 0.2865 \\\hline
\end{tabular}\\
Table 5. Simulated mean and MSE of the estimators $\tilde{\eta}$,
$\hat{\eta}$ and $\overset{\circ}{\eta}$.
\end{center}
Obviously, the MRE estimator $\overset{\circ}{\eta}$ always exhibit
the minimum squared error, as $\tilde\eta$ and $\hat\eta$ are
equivariant estimators of $\eta$. Notice also the biased character of
the maximum likelihood and MRE estimators. $\Box$ \end{myrema}

\begin{myrema}\rm
Although less interesting for the applications, let
us consider now the problem of estimating the scale parameter $\eta$
when the position parameter $\xi$ is known, say $\xi=\xi_0$. After
the shift $(y_1,\dots,y_n)\mapsto (y_1-\xi_0,\dots,y_n-\xi_0)$, the
statistical model remains invariant under the transformations
(dilations) of the form $(y_1,\dots,y_n)\mapsto (ay_1,\dots,ay_n)$,
for $a>0$. For the loss function
$W'_1(\eta,x)=(x-\nolinebreak\eta)^2/\eta^2$, the MRE estimator of
the scale parameter $\eta$ is
$$T_2=\frac{\Gamma(\frac{n+1}{2})}{\sqrt{2}\Gamma(\frac{n+2}{2})}\sqrt{\sum_{i=1}^n(Y_i-\xi_0)^2}=
\frac{B(\frac{n+1}{2},\frac12)}{\sqrt{2\pi}}
\sqrt{\sum_{i=1}^n(Y_i-\xi_0)^2}
$$
where $\Gamma$ and $B$ denote Euler's Gamma and Beta functions. In
fact, for the loss function $W'_1$, the MRE estimator of $\eta$ is
$$T_2(y_1,\dots,y_n)=\frac{\displaystyle\int_0^{\infty}v^nh_1(v(y_1-\xi_0),...,v(y_n-\xi_0))dv}
{\displaystyle\int_0^{\infty}v^{n+1}h_1(v(y_1-\xi_0),...,v(y_n-\xi_0))dv},$$
where
$$h_1(y_1,\dots,y_n)=\left(\frac{2}{\pi}\right)^{\frac{n}{2}}\exp
\left\{-\frac{1}{2}\sum_{i=1}^ny_i^2\right\}\cdot
I_{[0,+\infty[}(y_{1:n}).$$

To facilitate the notation, we suppose without loss of generality that
$\xi_0=0$. The change of variables
$t=\frac{1}{2}\sum_{i=1}^ny_i^2v^2$ shows that, for $k=n,n+1$,

\begin{gather*}\int_0^{\infty}v^kh_1(vy_1,...,vy_n)dv=
2^{\frac{n+k-1}{2}}\pi^{-\frac{n}{2}}\left(\sum_{i=1}^ny_i^2\right)^{-\frac{k+1}{2}}
\Gamma\left(\frac{k+1}{2}\right)I_{[0,+\infty[}(y_{1:n}),
\end{gather*}
and the assertion follows easily from this.

Note also that, when $\xi=\xi_0$,
$$\frac1n\sum_{i=1}^n(Y_i-\xi_0)^2$$ is
the minimum variance unbiased estimator of $\eta^2$. This is a
consequence of the Lehmann-Scheff\'{e} Theorem and the facts that
$\sum_{i=1}^n(Y_i-\xi_0)^2$ is a sufficient and complete statistic
and $\eta^{-2}\sum_{i=1}^n(Y_i-\xi_0)^2$ has distribution
$\chi^2(n)$. A little more work shows that
$$\frac{\Gamma(\frac{n}{2})}{\sqrt{2}\Gamma(\frac{n+1}{2})}\sqrt{\sum_{i=1}^n(Y_i-\xi_0)^2}=
\frac{B(\frac{n}{2},\frac12)}{\sqrt{2\pi}}
\sqrt{\sum_{i=1}^n(Y_i-\xi_0)^2}
$$
is the minimum variance unbiased estimator of $\eta$. $\Box$
\end{myrema}

%%\section{Acknowledgments}
%\begin{myack}\rm
%This work was supported by the Spanish Ministerio de Ciencia y
%Tecnolog\'ia under the project MTM2010-16845 and the Junta de
%Extremadura under the GR10064 grant.
%\end{myack}

%\section{References}

\end{document}